\documentclass{svjour2} 
\smartqed

\newcommand{\be}{\begin{equation}}    % for lazy typers
\newcommand{\ee}{\end{equation}}
\newcommand{\ba}{\begin{eqnarray}}
\newcommand{\ea}{\end{eqnarray}}

\catcode`@=11
\newcommand\eqalign[1]{\null\,\vcenter{\openup\jot\m@th
    \ialign{\strut\hfil$\displaystyle{##}$&$\displaystyle{{}##}$\hfil
        \crcr#1\crcr}}\,}

\begin{document}

\title{Quantitative predictions with detuned normal forms}
\author{Giuseppe Pucacco \and Dino Boccaletti  \and Cinzia Belmonte}
             
             \institute{G. Pucacco \at
              Physics Department, University of Rome ``Tor Vergata" and
INFN - Sez. Roma II\\
              \email{pucacco@roma2.infn.it} \and
              D. Boccaletti \at
              Mathematics Department, University of Rome ``La Sapienza" \\ \email{boccaletti@uniroma1.it}    \and
              C. Belmonte \at
              Physics Department, University of Rome ``La Sapienza" \\
              \email{ciaulalunae@fastwebnet.it}                 }

\date{Received: 10/12/2007 / Accepted: 15/04/2008}

\maketitle 

\begin{abstract}
The phase-space structure of two families of galactic potentials is approximated with a resonant detuned normal form. The normal form series is obtained by a Lie transform of the series expansion around the minimum of the original Hamiltonian. Attention is focused on the quantitative predictive ability of the normal form. We find analytical expressions for bifurcations of periodic orbits and compare them with other analytical approaches and with numerical results. The predictions are quite reliable even outside the convergence radius of the perturbation and we analyze this result using resummation techniques of asymptotic series.

\keywords{Normal forms of Hamiltonian systems \and Galactic potentials \and Stability of periodic orbits}
\end{abstract}

\section{Introduction}

Normal forms are of invaluable help in approximating the phase-space structure of non-integrable systems and in providing a picture of the dynamics in the regular domains (Giorgilli and Locatelli, 2006). In particular, {\it detuned} resonant normal forms allow us to investigate in detail several features of non-linear oscillators. Leaving free parameters, these tools have been extensively applied to get useful qualitative information (Cushman and Bates, 1997; Broer at al., 2003). However, {\it quantitative} predictions in concrete applications are as much as useful. To take a specific example, let us consider the bifurcations of periodic orbits: the stability--instability threshold in a suitable parameter space can be determined by studying the nature of critical points of either exact invariant functions in the normalizing variables or of approximate integrals in the original variables. The agreement between both methods is obtained through series expansions in characteristic parameters (Belmonte et al., 2007), for example the energy and the ellipticity of the equipotentials: in the case of the galactic `logarithmic' potential (Binney and Tremaine, 1987), the agreement with other analytical or numerical approaches is very good even truncating the normal form at a relatively low order. Attempts have also been made to treat perturbations of the isochronous sphere (Gerhard and Saha, 1991; Yanguas, 2001).

Aim of this work is to study in a more systematic way the quantitative predictive ability of these expansions. We investigate the result of different choices of the effective perturbation order of the detuning term. The usual approach to construct detuned normal forms is based on the idea of considering detuning zero-order terms as being of higher order and therefore treating them as part of the perturbation. In the process of normalization, new terms appear at orders (and with coefficients) that depend on the `effective' order of the detuning term. In turn, different terms appear in the process of inversion to construct approximate `third' integrals of motion (Contopoulos, 2004; Giorgilli, 2002). Benchmarks used to test the predictions are in all cases the outcome of other analytical approaches. The results point out that the best choice is always that of treating the detuning term as the lowest non-zero order term compatible with the symmetries of the problem at hand. 

A remarkable side-effect of expressing the stability--instability threshold as a series expansion, is that its predictive ability goes well beyond the radius of convergence of the perturbing expansion. In the `galactic' cases studied here, we show that the bifurcation energy of the major-axis periodic orbit is predicted with high accuracy up to values much higher than the `harmonic core' energy. We are using probably divergent series; however, their truncations stay close to actual functions. To justify these extended results, we try to estimate an optimal truncation by exploiting some resummation technique based on asymptotic series and continued fractions. 

The plan of the paper is as follows: in Section 2 we recall the procedure of detuned resonant normalization as applied to double-reflection symmetry potentials; in Section 3 we sketch the basics of the theory of stability of normal modes for these systems; in Section 4 we study the effective order of the detuning and compare the predictions with numerical data, analyzing the results using resummation techniques of asymptotic series; in Section 5 we give our conclusions.

\section{The Hamiltonian system}

We are interested in 2-degrees of freedom natural systems of the form
\be
   H({\bf p,q})= \frac{1}{2}(p_x^2+p_y^2) + V(x^2,y^2),
 \ee
with $V$ a smooth potential with an absolute minimum and reflection symmetry with respect to both axes. The motivation for the choice of this symmetry stems from its interest in problems of galactic dynamics. In particular, we will examine the following two examples
\ba
V_{L}&=&\frac12 \log(R^{2} + x^2 + y^2 / q^2),\label{v1}\\
V_{C}&=&\sqrt{R^{2} + x^2 + y^2 / q^2}-R.\label{v2}\ea
Both choices are useful models for elliptical galaxies: in particular, the logarithmic potential $V_{L}$ is used to model a core embedded in a dark matter halo. For every finite values of the ``core radius'' $R$, the choice $R=1$ can be done without any loss of generality. However, in both cases it is also of relevance the singular limit $R\rightarrow0$ associated to central density cusps. At high energy, the dynamics in the non-singular cases tend to the corresponding scale-free singular limits, respectively the singular logarithmic potential $V_{LS}$ and the ``conical'' potential $V_{CS}$. Since it is difficult to devise a perturbative approach to approximate the dynamics in the singular cases, one may hope to infer information on them from the study of the non-singular cases at sufficiently high values of the energy. 

With the choice $R=1$, the energy $E$ may take in both cases any non-negative value 
 \be 0 \le E < \infty. \ee
 The parameter $q$ gives the ``ellipticity'' of the figure and ranges in the interval
 \be\label{rangeq}
 0.6 \le q \le 1. \ee
 Lower values of $q$ can in principle be considered but correspond to an unphysical density distribution. Values greater than unity are included in the treatment by reversing the role of the coordinate axes. In order to implement the normalization algorithm, the Hamiltonian has to be expressed as a series expansion around the equilibrium. Performing the scaling transformation
\be
p_y \longrightarrow {\sqrt{q}} \, p_{y}, \quad  
y \longrightarrow \frac{y}{\sqrt{q}},\ee
the Hamiltonian can be expanded as 
\be
H = \sum_{k=0}^{\infty} H_{k} = \frac{1}{2}\biggl(p_x^2+ {{1} \over {q}} p_y^2\biggr) + \sum_{k=0}^{\infty} b_{k} s^{k+1},
\ee
where
\be
s = x^2 + {{1} \over {q}} y^2\ee
and $b_{k}$ are real coefficients (with $b_{0}=1/2$). The series expansions of the potentials (\ref{v1},\ref{v2}) around the origin, with $R=1$, are:
\be\label{ELP}
 V_{L} = \frac12 s -  \frac{1 }{2 \cdot 2}  s^{2} + \frac{1 }{2 \cdot 3}  s^{3} - \frac{1 }{2 \cdot 4}  s^{4} + \dots
 \ee
and
\be\label{E2P}
 V_{C} = \frac12 s -  \frac{1 \cdot 1}{2 \cdot 4} s^{2} + \frac{1 \cdot 1 \cdot 3}{2 \cdot 4 \cdot 6} s^{3} - \frac{1 \cdot 1 \cdot 3 \cdot 5}{2 \cdot 4 \cdot 6 \cdot 8} s^{4} + \dots
 \ee
respectively.

We look for a new Hamiltonian in the new canonical variables $\bf{P,Q}$, given by 
\begin{equation}\label{HK}
     K({\bf{P,Q}})=\sum_{n=0}^{\infty}K_n ({\bf{P,Q}}),
  \end{equation}
with the prescription that 
\be\label{NFD}
\{H_0,K\}=0.
\ee
The zero order (unperturbed) Hamiltonian,  
\be\label{Hzero}
H_{0} = \frac12 (P_X^2 + X^2) + \frac{1}{2q} (P_Y^2  +Y^2),
\ee
with unperturbed frequencies $\omega_1=1$ and $\omega_2 = 1/q$, is expressed in terms of the new variables found at each step of the normalizing transformation. In these and subsequent formulas we adopt the convention of labeling the first term in the expansion with the index zero: in general, the `zero order' terms are quadratic homogeneous polynomials and terms of {\it order} {\it n} are polynomials of degree $n+2$. 

It is customary to refer to the normal form constructed in this case as a ``Birkhoff'' normal form (Birkhoff, 1927). The presence of terms with small denominators in the expansion, forbids in general its convergence. It is therefore more effective to work since the start with a {\it resonant normal form} (Sanders \& Verhulst, 1985), which is still non-convergent, but has the advantage of avoiding the small divisors associated to a particular resonance. To catch the main features of the orbital structure, we therefore approximate the frequencies with a rational number plus a small ``detuning'' 
\be\label{DET}
\frac{\omega_1}{\omega_2} = q = \frac{m_{1}}{m_{2}} + \delta. \ee
We speak of a {\it detuned} ($m_{1}$:$m_{2}$) {\it resonance}, with $m_{1}+m_{2}$ the {\it order} of the resonance and, performing
the rescaling
\be\label{newE}{\cal H} := \frac{m_{2} H}{\omega_2} = m_{2} q H,\ee
we redefine the Hamiltonian in the form
\be\label{detH}
{\cal H} = \sum_{k=0}^{\infty} {\cal H}_{k} = \frac12 [m_{1} (p_x^2+x^2) + m_{2} (p_y^2+y^2)]+ 
m_{2} [ {\scriptstyle{\frac12}} \delta (p_x^2+x^2) + q\sum_{k=1}^{\infty} b_{k} s^{k+1} ].\ee 
The procedure is now that of an ordinary resonant ``Birkhoff--Gustavson'' normalization (Gustavson, 1966; Moser, 1968) with two variants: the coordinate transformations are performed through the Lie transform and the detuning quadratic term is treated as a term of higher order and put in the perturbation.

The new coordinates $\bf{P,Q}$ result from the canonical transformation
  \be\label{TNFD}
  ({\bf{p,q}}) = T_{\chi} (\bf{P,Q}).\ee
Considering a generating function $\chi$, the Lie transform operator $T_{\chi}$ is defined by (Boccaletti and Pucacco, 1999)
\begin{equation}\label{eqn:OperD-F}
    T_{\chi} \equiv \sum_{k=0}^{\infty} M_k
\end{equation}
where
\be M_0 = 1, \quad M_k = \sum_{j=1}^k \frac{j}{k} L_{\chi_j} M_{k-j}.\ee
The functions $\chi_k$ are the coefficients in the expansion of the generating function of the canonical transformation and the linear differential operator $L_{g}$ is defined through the Poisson bracket, $L_{g}(\cdot)=\{g,\cdot\}$.

The terms in the hew Hamiltonian are determined through the recursive set of linear partial differential equations (Giorgilli, 2002)
\be\label{EHK} \begin{array}{ll}
    &K_0={\cal H}_0 ,\\ \\
    &K_1={\cal H}_1+M_{1}{\cal H}_0 =  {\cal H}_1+L_{\chi_1}{\cal H}_0,\\ \\
    &K_2={\cal H}_2+M_{1}{\cal H}_1+M_{2}{\cal H}_0 =  
               {\cal H}_2+L_{\chi_1}{\cal H}_1 + \frac12 L^{2}_{\chi_1}{\cal H}_0+L_{\chi_2}{\cal H}_0 ,\\ \\
    &\quad \;\; \vdots \\ \\
    &K_n= {\cal H}_n  +\sum_{j=1}^{n-1}M_{n-j}{\cal H}_j +M_n {\cal H}_0,
\end{array}\ee
where $L^{2}_{g}(\cdot)=\{g,\{g,\cdot\}\}$. `Solving' the equation at the $n$-th step consists of a twofold task: to find $K_{n}$ {\it and} $\chi_n$. The unperturbed part of the Hamiltonian, ${\cal H}_0$, determines the specific form of the transformation. In fact, the new Hamiltonian $K$ is said to be {\it in normal form} if, analogously to (\ref{NFD}),
\be
\{{\cal H}_0,K\}=0,
\ee
 is satisfied. We observe that, in view of the reflection symmetries of the potentials (\ref{v1},\ref{v2}), in the chain  (\ref{EHK}) they appear only terms with even index and so the normal form itself is composed by even index terms only.

We have to discuss how to treat the detuning term: it is considered as a higher order term and the most natural choice is to put it into ${\cal H}_2$. However, there is no strict rule for this and one may ask which is the most `useful' choice, always considering that applications are based on series expansions with coefficients depending on $q$. We remark that, different choices of the {\it effective order}, say $d$, of the detuning, lead to different terms of higher order in the normal form. We also observe that, whatever the choice made, the algorithm devised to treat, step by step, the system (\ref{EHK}) must be suitably adapted to manage with polynomials of several different orders. In practice, since at each step the actual order of terms associated to detuning is lower than the corresponding effective order, the algorithm is adapted by incorporating routines already used at previous steps. In practice, at step say $j$, we have an equation of the form
\be
K_j = {\cal H}_j + A_j + \delta B_{j-d} + \delta^2 B_{j-2d} + ... + L_{\chi_j} {\cal H}_0,\ee
where $A_i, B_i$ are homogeneous polynomials of degree $i+2$ coming from previous steps. As usual, the algorithm is designed to identify in all terms with the exclusion of 
\be
L_{\chi_j} {\cal H}_0 \equiv -L_{{\cal H}_0} (\chi_j),\ee
monomials in the kernel of the linear operator $L_{{\cal H}_0}$. These monomials are used to construct $K_j$: the remaining terms are used to find $ \chi_j $ in the standard way. It is clear that both the normal form and the generating function are affected by the effective order of the detuning term.

In both cases (\ref{ELP},\ref{E2P}) investigated here, with the detuning treated as a term of order 2, the next appearance of a related term is in $K_6$. Rather, if it is treated as a term of order 4, the next appearance of a related term is in $K_8$. Truncating at order 6 (polynomials of degree 8) is therefore sufficient to make a comparison with other predictions not sensitive to the detuning.

\section{Stability of periodic orbits}

A very useful setting in which to perform quantitative predictions is that of the stability threshold of normal modes and/or periodic orbits in general position (de Zeeuw and Merritt, 1983; Fridman and Merritt, 1997). In Belmonte et al. (2007) we have shown that these predictions can be obtained both exploiting the nature of critical points of the normal form constrained on the manifold determined by the first integral and studying fixed points on a surface of section constructed using the approximate integral in the original coordinates. The agreement between the two methods is up to the level of the truncation if the transition curve in the parameter space is expressed as a series expansion around the given resonance. The choice of the method can be done simply on the basis of computational simplicity: for example, if one is interested in the instability transition of the normal modes, to work with the normal form is usually easier. Moreover, it is simpler to use `action-angle--{\it like}' variables, defined through the transformation
\ba\label{AAV}
X &=& \sqrt{2 J_1} \cos \theta_1,\quad
P_X = \sqrt{2 J_1} \sin \theta_1,\\
Y &=& \sqrt{2 J_2} \cos \theta_2,\quad
P_Y = \sqrt{2 J_2} \sin \theta_2.\ea
A case that is both representative of the state of affairs and useful in galactic applications is that of the stability of the  {\it x}-axis periodic orbit (the `major-axis orbit', if $q$ is in the range (\ref{rangeq})). Among possible bifurcations from it, the most prominent is usually that due to the 1:2 resonance, producing the `banana' and `anti-banana' orbits (Miralda-Escud\'e and Schwarzschild, 1989). We will mostly investigate this problem in the potentials  (\ref{v1},\ref{v2}) and will briefly discuss other less relevant cases.

Let us consider the 1:2-symmetric resonant normal form, so that the frequency ratio (\ref{DET}) now is
\be\label{DET12}
\omega_1/\omega_2 = 1/2 + \delta.\ee
The lowest order incorporating the resonance is 4 and we can write the corresponding terms as
\ba\label{NF12}
K_0 &=&{\cal H}_{0} = J_1 +  2 J_2 ,\\
K_2 &=&2 \delta J_1 - P^{(2)}(J_1 , J_2),\\
K_4 &=&P^{(3)} (J_1 , J_2) + k J_1^2 J_2 \cos(4 \theta_1 - 2 \theta_2).
\ea
where the polynomials $P^{(2)}$ and $P^{(3)}$ are homogeneous of degree $2$ and $3$. In the logarithmic and conical case they are respectively
\begin{eqnarray*}
P^{(2)}_{L} &=& \frac34 \left(q J_{1}^2 + \frac1{q} J_{2}^2 \right) + J_{1} J_{2},    \\
P^{(3)}_{L} &=& q \left(\frac56 - 
\frac{17}{16} q \right) J_{1}^3 + \left(\frac{13}{12} - 
\frac32 q \right) J_{1}^2 J_{2} - \left(\frac5{12} - 
\frac{3}{4q} \right) J_{1} J_{2}^2 + 
\frac{29}{96 q^{2}} J_{2}^3
\end{eqnarray*}
and
\begin{eqnarray*}
P^{(2)}_{C} &=& 
\frac38 \left(q J_{1}^2 + \frac1{q} J_{2}^2 \right) + \frac12 J_{1} J_{2},    \\
P^{(3)}_{C} &=& q \left(\frac{5}{16} - \frac{17}{64} q \right) J_{1}^3 + 
\left(\frac{11}{24} - \frac38 q \right) J_{1}^2 J_{2} - 
\left(\frac5{48} - \frac{3}{8q} \right) J_{1} J_{2}^2 + 
\frac{23}{128 q^{2}} J_{2}^3.
\end{eqnarray*}
The constant $k$ in front of the resonant term in $K_4$ is respectively $q/8$ in the logarithmic and $(1+q)/32$ in the conical case. For simplicity we have included the detuning term, $2 \delta J_1$, in $K_2$. The generating functions have, in both cases, $\chi_{0}=1$ and $\chi_{1}=0$. The first non trivial term is
\begin{eqnarray*}
\chi_{2} &=&- \frac12 J_1 (q J_2 + J_1) \sin 2 \theta_1 - \frac{q}{16} J_1^2 \sin 4 \theta_1 +\\
             &&- \frac14 J_2 \left(J_1+\frac{J_2}{q} \right) \sin 2 \theta_2 -\frac{1}{32q} J_2^2 \sin 4 \theta_2 +\\
             &&+ \frac14 J_1J_2 \left( \sin(2 \theta_1 - 2 \theta_2) - \frac13 \sin(2 \theta_1 + 2 \theta_2) \right).
\end{eqnarray*}
in the logarithmic case and $2 \chi_{2}$ in the conical case. 

Action-angle--like variables are singular on axial orbits. We have to use mixed variables: action--angle variables on the normal mode and Cartesian variables on the normal bundle to it. To analyze the stability, we determine the condition for the normal mode to be a critical curve of the Hamiltonian in these coordinates and assess its nature by considering the function
\be\label{LM1}
K^{(\mu)}=K+\mu {\cal H}_0,\ee
where $\mu$ has to be considered as a {\it Lagrange multiplier} to take into account that there is the constraint 
\be {\cal H}_0 = {\cal E}\ee 
associated to the existence of the second integral. The Lagrange multiplier is found by imposing
\be\label{LM2}
{\rm d} \, K^{(\mu)} =0,\ee
that is the total differential of (\ref{LM1}) vanishes on the normal mode. Its nature is assessed by computing the matrix of second derivatives of $K^{(\mu)}$: if the Hessian determinant of the second variation is positive definite, the mode is elliptic stable; if it is negative definite, the mode is hyperbolic unstable (Kummer, 1977; Contopoulos, 1978). 

The equation ${\rm det}[{\rm d}^{2}K^{(\mu)}({\cal E})]=0$ is an algebraic equation of degree $ M $ in ${\cal E}$, where $M$ is the order of truncation of the normal form. At order $M=4$ it is 
\ba
\label{d12L}
&& \left(48 - 96 q + 24 (3 q -1) {\cal E} + (26 - 153 q + 153  q^2) {\cal E}^2 \right) 
\nonumber \\
&& \times \left(48 - 96 q + 24 (3 q -1) {\cal E} + (26 - 159 q + 153  q^2) {\cal E}^2 \right)= 0\ea
in the logarithmic case (Belmonte et al. 2007) and, analogously, it can be shown to be 
\ba
\label{d12C}
&& \left(192 - 384 q + 48 (3 q -1) {\cal E} + (41 - 213 q + 153  q^2) {\cal E}^2 \right) \nonumber \\
&& \times \left(192 - 384 q + 48 (3 q -1) {\cal E} + (41 - 219 q + 153  q^2) {\cal E}^2 \right)= 0\ea
in the conical case.

The solutions of these equations give a set of curves in the $q-{\cal E}$ plane. These are the loci of transition to instability of the normal mode and of bifurcation of a new family: among the set, the meaningful ones in the present case are those corresponding to the detuned 1:2 resonance. Therefore, the relevant solutions are those possessing a leading order going as
\be
{\cal E}_{\rm crit}(q) \sim q - \scriptstyle\frac12 \ee
and this suggests to expand the selected solution as a power series in the detuning. However, in order to get a form usable in comparison with other results (for example coming from a numerical treatment) it is necessary to use a `physical' energy variable rather than the parameter ${\cal E}$. The conversion is possible if the physical energy $E$ appears explicitly. According to the rescaling (\ref{newE}), we assume that $ m_{2} q E $ is the constant `energy' value assumed by the truncated Hamiltonian $K$. In the present instance $m_{2}=2$ so that, on the {\it x}-axis orbit, the new Hamiltonian is a series of the form
\be\label{KIa}
K = 2 q {\cal E} - \frac34 q {\cal E} ^{2}+ ... = 2qE.\ee
The series (\ref{KIa}) can be inverted to give
\be
{\cal E}= E + \frac38 E^2 + ...\ee
and this can be used in the treatment of stability to replace ${\cal E}$ with $E$. 
The solutions can therefore be expressed as
\be\label{sqe}
E_{\rm crit}(q) = \sum_{k=1}^{M/2} c_{k}{\left(q-\frac12 \right)}^{k}
\ee
and in this form they can be used for quantitative predictions. The reason for a truncation order $M/2$ for this series is due to the fact that we now work with the energy rather than with phase-space coordinates: since $M$ is even, the order is always an integer. A similar procedure can be followed for other bifurcations, keeping in mind that the lowest order to be included in the normal form in order to capture an $m_{1}$:$m_{2}$ symmetric resonance is $2\times(m_{1}+m_{2}-1)$ (Tuwankotta and Verhulst, 2000). Below, we will briefly touch on the case of the 1:1 resonance leading to the bifurcation of the {\it loop} orbit.

\section{Results and comparison with other methods}

In Belmonte et al. (2007) the series (\ref{sqe}) for the stability-instability transition of the resonant periodic orbits from normal modes have been computed for the potential (\ref{ELP}) up to order $M=6$. Here we extend those results to higher orders and to the other case of potential (\ref{E2P}). 

 \begin{table}
 \centering
    \begin{tabular}{||c||c|c||c|c||}
      \hline
      $$ & \multicolumn{2}{c||} {{\rm Potential} $V_{L}$} & \multicolumn{2}{c||} {{\rm Potential} $V_{C}$} \\ \hline
 $k$ & ${\rm Banana}$ &      ${\rm Anti-banana}$ & $ {\rm Banana} $  &   $ {\rm Anti-banana} $ \\
									   \hline 
& & 	& &								   \\
 $1$ & $ 8 $  &   $ 8 $ & $ 16 $  &   $ 16 $ \\ 
& &  & & \\
 $2$ & $ -{{20}\over{3}} $  &   $ {{28}\over{3}} $ & $ {{248}\over{3}} $  &   $ {{536}\over{3}} $\\ 
 & &  & & \\
 $3$ & $ \frac{268}{9} $  &   $ \frac{460}{9} $ & $ \frac{3608}{9} $  &   $ \frac{18584}{9} $\\
 & &  & & \\
 $4$ & $ -\frac{1724}{27} $  &   $ \frac{3928}{27} $ & $ \frac{43328}{27} $  &   $ \frac{657848}{27} $ \\
 & &  & & \\
 $5$ & $ \frac{79184}{405} $  &   $ \frac{267404}{405} $ 
        & $ \frac{525704}{81} $  &   $ \frac{23668304}{81} $\\
 & & & & \\
 $6$ & $ -\frac{567178}{1215} $  &   $ -\frac{510200857}{405} $ 
        & $  \frac{28118794}{1215} $  &   $ \frac{4304374384}{1215} $ \\
 & & & & \\
 $7$ & $ -{{30991946}\over{25515}} $  &   $ {{615376795556}\over{8505}} $ 
        & $  {{309430864}\over{3645}} $  &   $ {{31575390356}\over{729}} $ \\
        & & & & \\
 \hline
\end{tabular}
  \caption{Coefficients in the expansion (\ref{sqe}) with $M=14$ for the logarithmic potential (banana, 2nd column and anti-banana, 3rd column) and the conical potential (banana, 4th column and anti-banana, 5th column).}
\label{tl12}
\end{table}

\subsection{Choice of the effective order of the detuning}

In Table \ref{tl12}. we list the coefficients of (\ref{sqe}) giving the bifurcation of the 1:2 resonant periodic orbits (`banana' and `anti-banana') for the logarithmic potential (\ref{ELP}) and the conical potential  (\ref{E2P}). They have been obtained with a normal form truncated at order $M=14$ and with the detuning treated as a term of order 2. There is a complete agreement with the analytical approach based on the Poincar\`e-Lindstedt method (Scuflaire, 1995) and, as discussed below, there is a striking agreement with the numerical approach based on the Floquet method. 

The results obtained by Scuflaire (1995) are based on treating the transition as a parametric instability phenomenon and are therefore rooted in a quite different theoretical framework. He gets series relating the critical $q$ to the amplitude on the axial orbit: once translated to the form (\ref{sqe}), the agreement of all fractional coefficients is complete up to the order we have arrived at, $M/2 = 7$. On the other hand, if the detuning is treated as a term of order $d = 4$ or greater, we get a disagreement in the coefficients starting from $c_{3}$. This result confirms the analysis made above on the `propagation' of the detuned terms in the normal form and show that the choice $d=2$ is the optimal one. This choice has therefore been adopted for all subsequent predictions, which are compared with the numerical ones with a better agreement as long as the truncation order is increased.

What is remarkable in all these results is that the numerical analysis are performed with the {\it exact} logarithmic (or conical) potentials (\ref{v1},\ref{v2}), whereas the analytical predictions are, in any case, produced with methods based on the series expansions (\ref{ELP},\ref{E2P}) of the potentials with limited convergence radii. The usual attitude in normal form theory is to use the formal series as {\it asymptotic series} (Verhulst, 1996; Contopoulos et al., 2003; Efthymiopoulos et al., 2004) useful to approximate the dynamics in regular regions of phase space. We have adopted the same attitude in evaluating the reliability of the series giving the instability thresholds.

\subsection{Asymptotic series and resummation techniques}

Suppose we want to approximate a function $f(x)$ with an infinite power series and define the ``error'' by truncating the series at order $N$:
\be
\epsilon_{N}(|x-x_{0}|) = f(x) - \sum_{n=0}^{N} c_{n}(x-x_{0})^{n}.\ee
We say that the series is asymptotic to the function if
\be
\epsilon_{N}(|x-x_{0}|) \ll (x-x_{0})^{N}, \quad x \to x_{0}, \; N \; {\rm fixed}.\ee
Typically the terms in the series get smaller for a while, but eventually they start to increase (the series diverge!).
Since $c_{N+1}(x-x_{0})^{N+1}$ is an estimate of the {\it error}, we can find the {\it optimal order of truncation} $N=N_{\rm opt}$ determining the smallest term (Bender \& Orszag, 1978). The optimal order depends on the interval $|x-x_{0}|$: the larger the interval, the smaller $N_{\rm opt}$ and the accuracy in the approximation. On the contrary, for a convergent series, for every $x$ within the interval $|x-x_{0}|$
\be
\epsilon_{N}(|x-x_{0}|) = \sum_{n=N+1}^{\infty}c_{n} (x-x_{0})^{n} \to 0, \quad N \to \infty .\ee

Once reached the optimal order, it can be disappointing to discard terms coming from a costly high-order computation. There are however other sophisticated rules for `summing' divergent series which make use of all terms (Bender \& Orszag, 1978), like the construction of Pad\`e approximant. A related approach is that of constructing {\it continued fractions} from the original power series in the form
\be\label{CF}
f(x) = \frac{a_{0}}{1+\frac{a_{1} (x-x_{0})}{1+\frac{a_{2} (x-x_{0})}{1+...}}}.\ee
The coefficients $a_{n}$ can be computed from the $c_{n}$ of the original series by expanding the continued fraction, comparing it with the original series and equating terms of the same order. 
Successive approximants obtained by truncating the fraction at various order may give an improvement in the asymptotic convergence with respect to the original series (Khovanskii, 1963). In the following subsections we exploit these techniques to try to evaluate the optimal order of the series giving the instability thresholds.

 \begin{table}
 \centering
    \begin{tabular}{||c||c|c|c|c||}
      \hline
      $$ & \multicolumn{4}{c||} {$q$} \\ \hline
 $N$ & $0.6$ &      $0.7$ & $ 0.8 $  &   $ 0.9 $ \\
									   \hline 
& & 	& &								   \\
 $1$ & $ 0.800000 $  &   $ 1.60000 $ & $ 2.40000 $  &   $ 3.20000 $ \\
& & 	& &								   \\
 $2$ & $ 0.733333 $  &   $ 1.33333 $ & $ 1.80000 $  &   $ 2.13333 $ \\
& & 	& &								   \\
 $3$ & $ 0.763111 $  &   $ 1.57156 $ & $ 2.60400 $  &   $ 4.03911 $ \\
 & & 	& &								   \\
 $4$ & $ 0.756726 $  &   $ 1.46939 $ & $ 2.08680 $  &   $ --- $ \\
 & & 	& &								   \\
 $5$ & $ 0.758681 $  &   $ 1.53196 $ & $ 2.56190 $  &   $ --- $ \\
 & & 	& &								   \\
 $6$ & $ 0.758214 $  &   $ 1.50208 $ & $ 2.22160 $  &   $ --- $ \\
 & & 	& &								   \\
 $7$ & $ 0.758336 $  &   $ 1.51763 $ & $ 2.48724 $  &   $ --- $ \\
        & & & & \\ 
        \hline
        $E_{B}$ & $ 0.758 $  &   $ 1.513 $ & $ 2.401 $  &   $ 3.646 $ \\
 \hline
\end{tabular}
  \caption{Subsequent truncations of expansion (\ref{sqe}) with $M=14$ for the logarithmic potential (banana). $E_{B}$ is the value obtained by means of the Floquet method.}
\label{pl12}
\end{table}

\subsection{Bifurcation in the logarithmic potential}

We start by investigating the bifurcation of the 1:2 resonant periodic orbits (`banana' and `anti-banana') from the {\it x}-axis orbit of the logarithmic potential. Truncating to order 14, it is possible to get the loci in the $q-E$ plane as in (\ref{sqe}) with the coefficients of Table \ref{tl12}. To evaluate these predictions, we treat the series (\ref{sqe}) as an asymptotic series and evaluate  truncations by computing the successive partial sums
\be
E_{N}(q) = \sum_{k=1}^{N} c_{k}{\left(q-\frac12 \right)}^{k}, \quad N=1,...,M/2.
\ee
In Table \ref{pl12}. we report these partial sums for the banana, with $q$ in the range from $0.6$ to $0.9$ and compare them with the numerical values obtained by means of the Floquet method (Miralda-Escud\'e and Schwarzschild, 1989; Belmonte et al. 2007) given in the last row. The numerical values of the partial sums are given with 6 digits just to show more clearly the asymptotic behaviour: we can see that, up to $q=0.8$, the predictions are apparently still (slowly) converging at $N=7$. Only at the rather extreme value $q=0.9$ we get an `optimal' truncation order $N_{\rm opt} = 3$, with a $10 \%$ error on the exact value of the critical energy.

We may wonder if the continued fraction may help in speeding up the convergence rate: that actually this is the case can be seen in Table \ref{cfl12}. where we report the partial sums computed with a continued fraction (\ref{CF}) truncated at $N \in [1,7]$. For all values of $q$ up to $0.8$, $N=6$ {\it is enough to reach a precision comparable to the numerical error}. For $q=0.9$ we get an optimal truncation order $N_{\rm opt} = 5$, with a $3 \%$ error on the exact value of the critical energy.

 \begin{table}
 \centering
    \begin{tabular}{||c||c|c|c|c||}
      \hline
      $$ & \multicolumn{4}{c||} {$q$} \\ \hline
 $N$ & $0.6$ &      $0.7$ & $ 0.8 $  &   $ 0.9 $ \\
									   \hline 
& & 	& &								   \\
 $1$ & $ 0.800000 $  &   $ 1.60000 $ & $ 2.40000 $  &   $ 3.20000 $ \\
& & 	& &								   \\
 $2$ & $ 0.738462 $  &   $ 1.37143 $ & $ 1.92000 $  &   $ 2.40000 $ \\
& & 	& &								   \\
 $3$ & $ 0.753917 $  &   $ 1.45915 $ & $ 2.14359 $  &   $ 2.81722 $ \\
 & & 	& &								   \\
 $4$ & $ 0.758120 $  &   $ 1.50678 $ & $ 2.33218 $  &   $ 3.31985 $ \\
 & & 	& &								   \\
 $5$ & $ 0.758295 $  &   $ 1.51190 $ & $ 2.37192 $  &   $ 3.51293 $ \\
 & & 	& &								   \\
 $6$ & $ 0.758323 $  &   $ 1.51399 $ & $ 2.40779 $  &   $ --- $ \\
 & & 	& &								   \\
 $7$ & $ 0.758313 $  &   $ 1.51281 $ & $ 2.38179 $  &   $ --- $ \\
        & & & & \\ 
        \hline
        $E_{B}$ & $ 0.758 $  &   $ 1.513 $ & $ 2.401 $  &   $ 3.646 $ \\
 \hline
\end{tabular}
  \caption{Subsequent truncations of the continued fraction (\ref{CF}) with $M=14$ for the logarithmic potential (banana).}
\label{cfl12}
\end{table}

Another check is that based on the more uncertain anti-banana transition, for which reliable numerical predictions exist only in the range $0.6 \div 0.7$. We put the results together in the same Table \ref{ab12}. Even here, the continued fraction is clearly more efficient. 

 \begin{table}
 \centering
    \begin{tabular}{||c||c|c|c|c||}
      \hline
      $$ & \multicolumn{4}{c||} {$q$} \\ \hline
 $N$ & $0.6$ &      $0.6$ & $ 0.7 $  &   $ 0.7 $ \\
									   \hline 
& & 	& &								   \\
 $1$ & $ 0.800000 $  &   $ 0.800000 $ & $ 1.60000 $  &   $ 1.60000 $ \\
& & 	& &								   \\
 $2$ & $ 0.893333 $  &   $ 0.905660 $ & $ 1.97333 $  &   $ 2.08696 $ \\
& & 	& &								   \\
 $3$ & $ 0.944444 $  &   $ 1.006320 $ & $ 2.38222 $  &   $ -2.32000 $ \\
 & & 	& &								   \\
 $4$ & $ 0.958993 $  &   $ 0.966571 $ & $ 2.61499 $  &   $ 3.09306 $ \\
 & & 	& &								   \\
 $5$ & $ 0.965595 $  &   $ 0.969092 $ & $ 2.82627 $  &   $ 3.38465 $ \\
 & & 	& &								   \\
 $6$ & $ 0.968026 $  &   $ 0.969675 $ & $ 2.98186 $  &   $ 3.59815 $ \\
 & & 	& &								   \\
 $7$ & $ 0.969089 $  &   $ 0.969877 $ & $ 3.11793 $  &   $ 4.15(*) $ \\
        & & & & \\ 
        \hline
        $E_{A}$ & $ 0.970 $  &   $ 0.970 $ & $ 4.292 $  &   $ 4.292 $ \\
 \hline
\end{tabular}
  \caption{Subsequent truncations of series (\ref{sqe}) (first and third columns) and of continued fraction (\ref{CF}) (second and fourth columns) with $M=14$ for the logarithmic potential (anti-banana: $E_{A}$ is the value obtained by means of the Floquet method.). The asterisk denotes a partial sum beyond the optimal truncation.}
\label{ab12}
\end{table}

Concerning the bifurcation of the {\it loop} orbit from the {\it y}-axis orbit, this happens in general at lower energies than that of the banana, therefore we presume to be predicted even better. This can be verified in the data of Table \ref{ll11}. where the continued fraction has been used. All partial sums are displayed since in any cases the optimal order is still unreached at $N=7$: when entries are repeated is because the rounding off hides a truncation error which is still decreasing.

 \begin{table}
 \centering
    \begin{tabular}{||c||c|c|c|c||}
      \hline
      $$ & \multicolumn{4}{c||} {$q$} \\ \hline
 $N$ & $0.6$ &      $0.7$ & $ 0.8 $  &   $ 0.9 $ \\
									   \hline 
& & 	& &								   \\
 $1$ & $ 0.80000 $  &   $ 0.600000 $ & $ 0.400000 $  &  $ 0.200000 $ \\
& & 	& &								   \\
 $2$ & $ 1.00000 $  &   $ 0.705882 $ & $ 0.444444 $  &   $ 0.210526 $ \\
& & 	& &								   \\
 $3$ & $ 1.04000 $  &   $ 0.720000 $ & $ 0.448000 $  &   $ 0.210909 $ \\
 & & 	& &								   \\
 $4$ & $ 1.03738 $  &   $ 0.719333 $ & $ 0.447891 $  &   $ 0.210903 $ \\
 & & 	& &								   \\
 $5$ & $ 1.03776 $  &   $ 0.719408 $ & $ 0.447900 $  &   $ 0.210904 $ \\
 & & 	& &								   \\
 $6$ & $ 1.03783 $  &   $ 0.719419 $ & $ 0.447901 $  &   $ 0.210904 $ \\
 & & 	& &								   \\
 $7$ & $ 1.03783 $  &   $ 0.719419 $ & $ 0.447901 $  &   $ 0.210904 $ \\
        & & & & \\ 
        \hline
        $E_{L}$ & $ 1.038 $  &   $ 0.719 $ & $ 0.448 $  &   $ 0.211 $ \\
 \hline
\end{tabular}
  \caption{Subsequent truncations of continued fraction (\ref{CF}) with $M=14$ for the logarithmic potential (loop). $E_{L}$ is the critical value obtained by means of the Floquet method.}
\label{ll11}
\end{table}

\subsection{Bifurcation in the conical potential}

In Table \ref{pc12}. we report the partial sums for the banana transition in potential (\ref{v2}), with $q$ in the range from $0.6$ to $0.7$ and compare them with the numerical values obtained by means of the Floquet method. In Table \ref{lc11}. are reported the data for the {\it loop} orbit: even here, all partial sums are displayed since in any cases the optimal order is still unreached at $N=7$: when entries are repeated is because the rounding off hides a truncation error which is still decreasing. 

In this potential, at higher values of the ellipticity, the bifurcation of the banana occurs, if any, at extremely high values of the energy: the same happens with the bifurcation of the loop at any $q$, with a critical energy much higher than in the logarithmic case. In both tables, it is clearly appreciable the accelerated convergence of the continued fraction up to those extreme values of the transition energy. 

 \begin{table}
 \centering
    \begin{tabular}{||c||c|c|c|c|c|c||}
      \hline
      $$ & \multicolumn{6}{c||} {$q$} \\ \hline
 $N$ & $0.6$           &      $0.6$        & $0.65$          &      $0.65$ & $ 0.7 $  &   $ 0.7 $ \\
									   \hline 
& & 	& &	& &							   \\
 $1$ & $ 1.60000 $  &   $ 1.60000 $ & $ 2.40000 $  &   $ 2.40000 $ & $ 3.20000 $  &   $ 3.20000 $ \\
& & 	& &	& &							   \\
 $2$ & $ 2.42667 $  &   $ 3.31034 $ & $ 4.26000 $  &   $ 10.6667 $ & $ 6.50667 $  &   $ -96.0000 $\\
& & 	& &	& &							   \\
 $3$ & $ 2.82756 $  &   $ 3.20501 $ & $ 5.61300 $  &   $ 9.22367 $ & $ 9.71378 $  &   $ 113.029 $\\
 & & 	& &	& &							   \\
 $4$ & $ 2.98803 $  &   $ 3.80153 $ & $ 6.42540 $  &   $ 13.0622 $ & $ 12.2814 $  &   $ -41.9262 $\\
 & & 	& &	& &							   \\
 $5$ & $ 3.05293 $  &   $ 3.09696 $ & $ 6.91825 $  &   $ 7.67725 $ & $ 14.3582 $  &   $ 23.1215 $\\
 & & 	& &	& &							   \\
 $6$ & $ 3.07607 $  &   $ 3.08855 $ & $ 7.18186 $  &   $ 7.45669 $ & $ 15.8394 $  &   $ 18.5519 $\\
 & & 	& &	& &							   \\
 $7$ & $ 3.08456 $  &   $ 3.08865 $ & $ 7.32691 $  &   $ 7.46075 $ & $ 16.9260 $  &   $ 18.6518 $\\
        & & & & & & \\ 
        \hline
        $E_{B}$ & $ 3.09 $  &   $ 3.09 $ & $ 7.46 $  &   $ 7.46 $ & $ 18.60 $  &   $ 18.60 $ \\
 \hline
\end{tabular}
  \caption{Subsequent truncations of series (\ref{sqe}) (first, third and fifth columns) and of continued fraction (\ref{CF}) (second, fourth and sixth columns) with $M=14$ for the conical potential (banana). }
\label{pc12}
\end{table}

 \begin{table}
 \centering
    \begin{tabular}{||c||c|c|c|c||}
      \hline
      $$ & \multicolumn{4}{c||} {$q$} \\ \hline
 $N$ & $0.6$ &      $0.7$ & $ 0.8 $  &   $ 0.9 $ \\
									   \hline 
& & 	& &								   \\
 $1$ & $ 1.60000 $  &   $ 1.20000 $ & $ 0.80000 $  &  $ 0.400000 $ \\
& & 	& &								   \\
 $2$ & $ \infty $       &   $ 4.80000 $ & $ 1.60000 $  &   $ 0.533333 $ \\
& & 	& &								   \\
 $3$ & $ 21.6000 $  &   $ 4.10323 $ & $ 1.54074 $  &   $ 0.529870 $ \\
 & & 	& &								   \\
 $4$ & $ 8.50000 $  &   $ 3.64608 $ & $ 1.51495 $  &   $ 0.529213 $ \\
 & & 	& &								   \\
 $5$ & $ 10.0922 $  &   $ 3.71312 $ & $ 1.51728 $  &   $ 0.529239 $ \\
 & & 	& &								   \\
 $6$ & $ 9.92646 $  &   $ 3.70788 $ & $ 1.51715 $  &   $ 0.529238 $ \\
 & & 	& &								   \\
 $7$ & $ 9.97972 $  &   $ 3.70929 $ & $ 1.51717 $  &   $ 0.529238 $ \\
        & & & & \\ 
        \hline
        $E_{L}$ & $ 10.38 $  &   $ 3.710 $ & $ 1.517 $  &   $ 0.529 $ \\
 \hline
\end{tabular}
  \caption{Subsequent truncations of continued fraction (\ref{CF}) with $M=14$ for the conical potential (loop).}
\label{lc11}
\end{table}

\subsection{Predictive ability of detuned normal forms}

The convergence radius of both expansions (\ref{ELP}--\ref{E2P}) in terms of the variable $s$ is of the order unity. The corresponding energy level is $E\sim0.35$ for the logarithmic potential $V_{L}$ and $E\sim0.4$ for the conical potential $V_{C}$. From the results presented above, it seems that the predictions obtained from the normal form are quite reliable up to energy levels much higher than these. The two potentials examined here show a common orbit structure: however, features corresponding to the same phenomenon happen to occur at higher energies for $V_{C}$ than for $V_{L}$. For example, from Tables \ref{cfl12} and \ref{pc12} we see that, for $q=0.6$ (namely, the ellipticity with the lowest values of the bifurcation energy), the banana bifurcates at $E=0.758$ in $V_{L}$ and $E=3.09$ in $V_{C}$, with an astounding convergence of the analytical predictions (in particular, with the continued fraction). We are led to speculate about the possibility of extending this predictive ability to general features of the systems. Moreover, we remark that at the energy levels reported here, both $V_{L}$ and $V_{C}$ slightly depart from the behavior of their scale-free counterparts. In view of the relevance of these results, it is of great interest the evaluation of an effective region of validity of the normal form. Therefore, we plan to extend the analysis to the study of the conservations of the approximate integrals of motion and to the explicit solutions of the equations of motion.

\section{Conclusions}

The results presented in this work provide a comprehensive setting to understand some stimulating but unsettled body of results obtained in previous studies.  

In Belmonte et al. (2006) we have started to investigate the stability of axial orbits in the logarithmic potential, using a normal form truncated to the first order incorporating the resonance: we got quite a satisfactory agreement with other analytical and numerical predictions but, among other things, we pointed out the troubles due to sensible differences between results coming either from the normal form itself (`final' normalizing variables) or from the approximate integral (`initial' physical variables). 

In Belmonte et al. (2007) we made a substantial step forward based on higher order normal forms: we were able to reconcile the results in the two different sets of variables by presenting the predictions as suitable power series in the {\it same} parameter space and, truncating at least at order $M=6$, we provided good {\it quantitative} predictions. 

In the present paper we have tried to answer the natural question about the limits of validity of those predictions. Adopting the same approach usually followed in exploiting approximate invariants constructed in a high order perturbation theory, we have shown how the power series representing the instability thresholds in the relevant parameter space can be interpreted as asymptotic series. As such, their truncated sums can be used to find an optimal truncation order and, in case, they can be resummed with the technique of the continued fraction, obtaining an improvement in the convergence rate. The generality of this setting allows us to conjecture that those results can be extended to arbitrary resonances and to periodic orbits in general position.

\end{document}